# Electronic Transport and Fermi Surface Topology of Zintl Phase Compound $SrZn_2Ge_2$

M. K. Hooda[1,2], A. Chakraborty[1,3], S. Roy[4], R. Swami[1], A. Agarwal[1], P. Mandal[5], S. N. Sarangi[6], D. Samal[6,7], V. P. S. Awana[8], and Z. Hossain[1]

[1] Department of Physics, Indian Institute of Technology, Kanpur-208016, India
[2] Department of Energy Science, Sungkyunkwan University, Suwon 16419, Republic of Korea
[3] Institute of Physics, Johannes Gutenberg Universität, Staudinger Weg 7, 55128 Mainz, Germany
[4] Vidyasagar Metropolitan College, 39, Sankar Ghosh Lane, Kolkata 700006, India
[5] Department of Condensed Matter and Material Physics, S. N. Bose National Centre for Basic Sciences, Block JD, Sector III, Salt Lake, Kolkata 700106, India
[6] Institute of Physics, Bhubaneswar, Bhubaneswar-751005, India
[7] Homi Bhabha National Institute, Anushakti Nagar, Mumbai 400085, India
[8] CSIR National Physical Laboratory, New Delhi, 110012, India

We report a comprehensive study on the electronic transport properties of $SrZn_2Ge_2$ single crystals. The electrical resistivity of the compound exhibits metallic behavior, following a $T^2$ dependence below 35 K, consistent with the Fermi liquid behavior. However, a notable deviation is observed from this behavior at lower temperatures as a pronounced resistivity plateau emerges below 10 K. This plateau is remarkably robust, and persists under the magnetic fields of up to 10 T. Both the transverse and longitudinal magnetoresistance exhibit a crossover at critical field $B^*$ from weak-field quadratic-like to high-field unsaturated linear field dependence at low temperatures ($T \leq 50$ K). Possible sources of linear magnetoresistance are discussed based on the Fermi surface topology, classical and quantum transport models. The Hall resistivity data establish $SrZn_2Ge_2$ as a multiband system with contributions from both the electrons and holes. The Hall coefficient is observed to decrease with increasing temperature and magnetic field, changing its sign from positive to negative. The negative Hall coefficient observed at low temperatures in high fields and at high temperatures over the entire field range suggests that the highly mobile electron charge carriers dominate the electronic transport. Our first-principles calculations show that nontrivial topological surface states exist in $SrZn_2Ge_2$ within the bulk gap along the Γ-M path. Notably, these surface states extend from the valence to conduction band with their number varying based on the Sr and Ge termination plane. The Fermi surface of the compound exhibits a distinct tetragonal petal-like structure, with one open and several closed surfaces. Overall, these findings offer crucial insights into the mechanisms underlying the electronic transport of the compound.

## 1. Introduction

Zintl phase compounds are well known for their exceptional thermoelectric properties, and structural stability over a wide range of temperatures [1,2]. Among them, the $AM_2X_2$ family (A = Alkali, alkaline-earth, and rare-earth elements; M = transition metal; X = group 13-16 elements) has drawn significant attention because of their several interesting properties, such as superconductivity (SC), colossal magnetoresistance (MR), metamagnetism, spin density wave, and topological phases such as field-induced topological Hall effect [3-13]. Interestingly, SC was commonly observed for compounds crystallizing in tetragonal $ThCr_2Si_2$-type structure (space group *I4/mmm*) [3-7]. For example, fully gapped *s*-wave SC was observed in $SrPd_2Ge_2$ at $T_c$ ~ 2.70 K [6]. SC found in $AFe_2As_2$ (A= K, Cs, Rb) compounds exhibits a highly unconventional nature [4,5,11]. In case of $RbFe_2As_2$, SC occurs in the nematic phase [5], whereas, in $KFe_2As_2$, superconducting order parameter has a complex muti-gap structure [11]. Notably, the former is determined by hole Fermi surfaces and can switch from being nodeless to having octet line nodes depending upon the location of Fermi surfaces within the Brillouin zones [11]. Multiple topological phases have been found in the magnetic $AM_2X_2$ family of compounds through experimental and theoretical investigations [12-17]. Density functional theory calculations revealed a single pair of Weyl nodes in $EuCd_2As_2$, which could be tuned by changing the canting angle [16]. Magnetotransport measurements further found the anomalous Hall effect and negative longitudinal MR, indicative of a chiral anomaly, supporting the Weyl node picture in $EuCd_2As_2$ [16]. The application of high magnetic fields was observed to suppress A-type antiferromagnetic ordering in isostructural $EuZn_2As_2$, giving rise to large negative in-plane MR [17]. Like $EuCd_2As_2$, $EuZn_2As_2$ was also reported to exhibit the nonlinear anomalous Hall effect but of a smaller magnitude due to relatively weak spin-orbit coupling (SOC) [17]. The existence of massless 2D Dirac fermions was confirmed in the antiferromagnetic compounds $BaFe_2As_2$ and $SrFe_2As_2$ through the field-dependent infrared studies and theoretical calculations [12].

In comparison to their magnetic counterparts, the nonmagnetic Zint phase $AM_2X_2$ family of topological phases has received the least attention. Recent theoretical investigations of alkaline-earth based Zintl phases have predicted a plethora of new topological candidate materials, such as $SrIn_2As_2$ (dual topological insulator), A(Ca,Sr,Ba)$Cd_2As_2$ (Dirac semimetals under pressure), $CaZn_2Bi_2$, and $CaCd_2Bi_2$ (topological crystalline insulators) [15,18-20]. The experimental authentication of topological phases in the case of nonmagnetic Zintl phases is realized only in a limited number of materials. $SrAl_2Si_2$ and $CaAl_2Si_2$ present a good examples of it, where nontrivial topology was confirmed from the Landau level fan diagram obtained through the Shubnikov-de Hass oscillations experiment [9,21].

Motivated by the remarkable potential of 122 family compounds to explore the topological physics and exotic electronic transport properties, which are primarily governed by their distinctive Fermi surface topology [5,11,12,21,22], our investigation focuses on a nonmagnetic Zintl phase material, $SrZn_2Ge_2$, from this family. The crystal structure and synthesis of polycrystalline $SrZn_2Ge_2$ are already reported in the literature [23]. Moreover, a study on the electronic structure and the physical properties of $SrZn_2Ge_2$ is absent in the literature.

In this paper, we report the electronic transport properties of single-crystalline $SrZn_2Ge_2$ by means of MR, Hall resistivity measurements, and electronic band structure calculations. We discuss the features of the electronic transport data in the context of conventional charge carriers, Dirac fermions, and the Fermi surface topology of the compound. Our study unveils the nontrivial topology and exotic transport properties of $SrZn_2Ge_2$ such as linear transverse and longitudinal MR, field- and temperature-induced sign reversal of the Hall coefficient.

## 2. Experimental details and methods

Single crystals of $SrZn_2Ge_2$ were prepared using Pb-flux reactions. The first batch of crystals was grown with the stoichiometric ratio Sr:Zn:Ge:Pb = 1:1:1:3. We also tried other stoichiometric compositions, such as 1:2:2:5, 1:2:2:7, and 1:2:2:10. Each approach yielded very shiny and thin flakes of single crystals. The reactions took place in vacuum-sealed quartz tubes, where the reactants were first loaded into an alumina crucible, and then the crucible was put into a quartz tube. The mixture was heated to 1050 ºC in 20 h and maintained at this temperature for 24 h to obtain a homogeneous solution. In the next step, the furnace was slowly cooled to 500 ºC at the rate of 3 ºC/h, and single crystals were isolated from Pb-flux using centrifuging process. We would further like to mention here that changing the centrifuge temperature from 500 to 650 ºC did not affect the quality of crystals. For the measurements, we have selected crystals with best stoichiometry from the last two batches, based on the Energy dispersive X-ray (EDX) spectroscopy data.

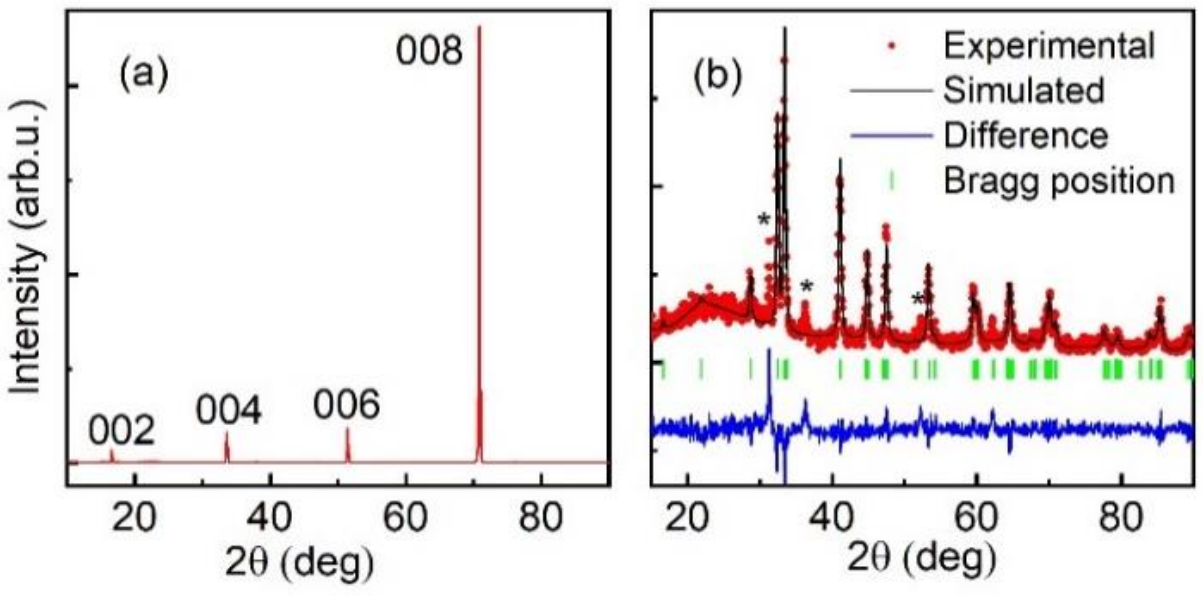

**Fig. 1.** X-ray diffraction pattern of $SrZn_2Ge_2$ (a) Single crystal. (b) Powder. The star marked peaks in the diffraction pattern are due to Pb-flux.

The X-ray diffraction (XRD) pattern obtained from a top surface of a representative $SrZn_2Ge_2$ single crystal is shown in Fig. 1. It shows the crystal growth along (00*l*) plane. The typical size of crystals is up to 3 mm in length, 2 mm in width, and 0.1 mm in thickness. EDX measurements carried out on several crystals showed a fairly homogeneous elemental distribution. The average atomic percentages of Sr (20.60%), Zn (41.06%), and Ge (38.33%) are close to the target stoichiometry (refer to the supplementary table S1 and Fig. S1). The Rietveld refinement performed on the powder XRD of several crushed single crystals showed that $SrZn_2Ge_2$ crystallizes in a tetragonal $ThCr_2Si_2$-type structure with space group *I4/mmm* (139). The lattice parameters obtained from the Rietveld refinement are $a = b = 4.391$ Å, and $c = 10.632$ Å, which are close to those reported in the literature [23].

Magnetotransport studies were performed in the temperature range of 2-300 K in magnetic fields up to 10 T using the Quantum Design Physical Property Measurement System and cryogen free measurement system from Cryogenic Limited. The experiments were done employing a standard four-probe technique.

## 3. Computational methods

The electronic structure calculations of bulk $SrZn_2Ge_2$, were performed using the density functional theory (DFT) in the plane wave basis set. We used the Perdew-Burke-Ernzerhof (PBE) [24] implementation of the generalized gradient approximation (GGA) for the exchange-correlation. This was combined with the projector augmented wave potentials [25,26] as implemented in the Vienna ab initio simulation package [27]. GGA calculations were carried out with and without SOC. In the former case, SOC was included as a second variational form to the original Hamiltonian. The kinetic energy cutoff of the plane wave basis set was chosen to be 500 eV. The momentum-space calculations for the Brillouin zone (BZ) integration of bulk were performed on a Γ- centered, 12 × 12 × 6 k-point grid. To calculate the surface spectral function for finite slab geometry of $SrZn_2Ge_2$, we construct the tight-binding model Hamiltonian by deploying atom-centered Wannier functions within the VASP2WANNIER90 [28]. Utilizing the obtained tight-binding model, we calculated the surface spectral function using the iterative Green's function method, as implemented in the Wannier Tools package [29].

## 4. Results and discussion

### *4.1. Electronic band structure and Fermi surface topology*

Bulk $SrZn_2Ge_2$ crystallizes in the tetragonal *I4/mmm* (139) space group and has two-fold ($C_{2i}$) and four-fold ($C_{4i}$) rotational symmetry, inversion, and five mirror symmetries, such as $m_{001}$, $m_{010}$, $m_{100}$, $m_{110}$, and $m_{1\text{-}10}$. Each conventional unit cell of $SrZn_2Ge_2$ consists of two formula units that are inverted with respect to the body center of the cell, as illustrated in Fig. 2(a). The $ZnGe_4$ tetrahedra accommodating Zn-Ge nearest neighbor bonds with a bond length of ~2.59 Å form an edge-shared network [see Fig. 2(b)] on the crystallographic *ab*-plane. The Sr atoms are situated in between the ZnGe planes, which are stacked along the *c* axis.

To comprehend the electronic band structure, we start with non-spin polarized calculations while enforcing spin degeneracy. We present the band dispersion along various high-symmetry directions and the density of states (DOS) of $SrZn_2Ge_2$ in Figs. 2(d) and 2(e). The finite contributions of DOS at the Fermi level suggest that the system is metallic. The four bands contributing to the DOS at the Fermi level are marked in green, red, purple, and cyan. The orbital projected DOS calculations show that the Ge-*p* states dominate the bands near the $E_F$, with Sr-*d*, Zn-*s*, and Zn-*p* states making small but significant contributions. The electronic band structure shows a large hole pocket in the top valence band (red) along the Γ-X path and a small hole pocket at the Z point, which has contributions from the two highest valence bands (shown in red and green).

The two lowest conduction bands generate two electron pockets at the A point and between the R-Z path, as shown in Fig. 2(d). Additionally, a very small electron pocket contributed by the first conduction band is observed near the Γ point. The presence of several holes and electron pockets near the $E_F$ suggests that both types of charge carriers are involved in the transport properties of the compound. To examine the impact of SOC on the low-energy electronic bands, we have also calculated energy dispersion within the GGA+SOC framework, as shown in Fig. 2(e). We find that SOC has a negligible effect on the electronic structure, except for some degeneracy lifting near the Γ point, as shown by the shaded orange region.

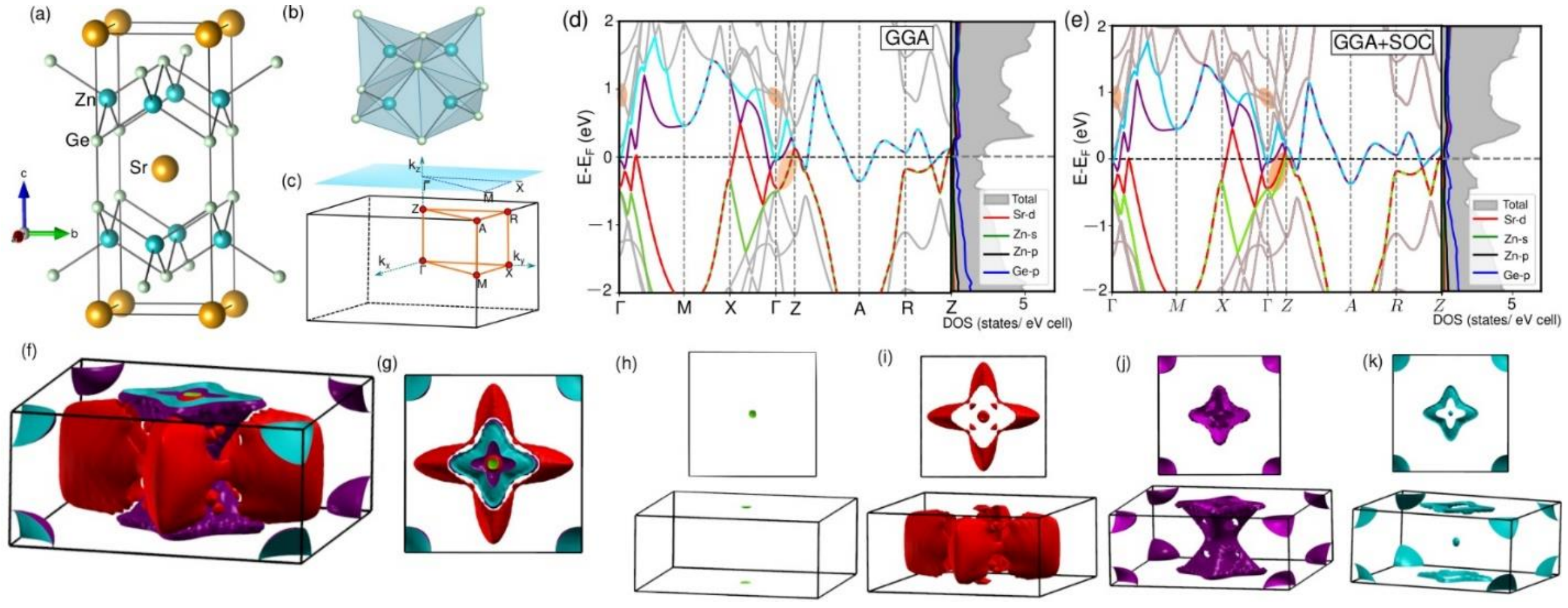

**Fig. 2.** (a) The conventional unit cell containing two formula units of bulk $SrZn_2Ge_2$. (b) The edge shared local $ZnGe_4$ tetrahedral network. (c) The irreducible Brillouin zone (BZ) of the bulk along with (001) projected surface as shown in cyan color. The density of states (grey shaded region) and bulk energy dispersion within (d) GGA and (e) GGA+SOC framework plotted along the high symmetry paths. The orbital decomposed DOS of Sr-*d* (red), Zn-*s* (green), Zn-*p* (black) and Ge-*p* (blue) suggest dominant contribution of Ge-*p* state near the Fermi energy. The SOC-induced lifted degeneracy of bands is shown in the orange shaded region. The side (f) and top (g) view of the overall Fermi surface depict a tetragonal petal like feature. The individual Fermi surface contributions of four bands (green, red, purple and cyan) near the Fermi level are presented in (h), (i), (j) and (k), respectively. The Fermi surface of the first conduction band (purple) is an open Fermi surface.

Next, we calculated the bulk Fermi surface from the four bands near the Fermi energy. We present the side and top view of the overall bulk Fermi surface of $SrZn_2Ge_2$ in Fig. 2(f) and Fig. 2(g), respectively. The figures exhibit a distinct tetragonal petal-like structure of the Fermi surface. The individual Fermi surfaces for the green, red, purple, and cyan bands are plotted in Figs. 2(h), 2(i), 2(j), and 2(k), respectively. We find closed Fermi surfaces for all the bands except the first conduction band as shown in purple in Fig. 2(d). The latter is an open Fermi surface with a hyperboloid shape that extends over the adjacent BZ.

### 4.2. Non-trivial surface states

Our examination of the bulk electronic structure, which have several Dirac band crossings near the Fermi level, led us to investigate further the topologically protected surface states near the Fermi energy. The zoomed in view of Dirac band crossings near the Fermi energy is shown in supplementary Fig. S2. We explore the existence of nontrivial surface states in the system using the layered stacking crystal structure. We calculated the spectral functions by creating a finite slab geometry along the (001) plane. The resulting bulk and surface spectral functions for Ge and Sr terminations are shown in Fig. 3(a) and 3(b), respectively.

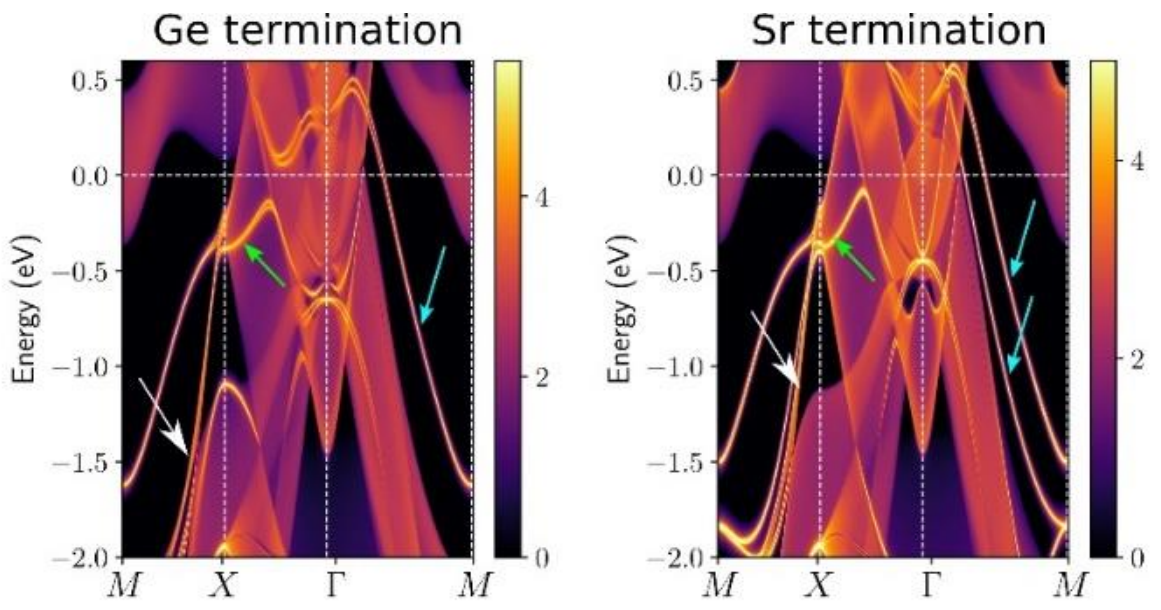

**Fig. 3.** The calculated momentum resolved spectral density plots along the X-Γ-M direction, (a) for Ge and (b) for Sr terminations. There exist non-trivial surface states within the bulk band gap along the Γ-M and M-X directions. The number of these non-trivial surface states (marked by cyan and white arrows) crucially depends on the type of the termination. The surface states marked with white arrows connect the valence and conduction sectors.

We found that along the Γ-X path, the surface states, as marked by green arrows, are deep within the bulk states. Interestingly, along the Γ-M path, there exist non-trivial surface states within the bulk band gap. The nontrivial surface states along the M-X path extend from the conduction to the valence bands as marked by the white arrows. For Ge and Sr termination, we find one and two surface bands (pointed with cyan arrows) in the bulk energy gap, respectively. Our findings demonstrate that the number of non-trivial surface states within the gap depends upon the termination plane, a critical aspect in comprehending this system.

### 4.3. *Electrical resistivity*

Figure 4 presents the temperature ($T$) dependence of zero-field electrical resistivity ($\rho$) measured within the $ab$-plane of two $SrZn_2Ge_2$ single crystals (marked by C1 and C2), which were grown in two different batches. Both crystals show almost similar behavior of $\rho(T)$, except crystal C1 exhibits an incipient superconducting drop at 7.2 K due to Pb flux residue [30]. At room temperature, $\rho$ value for crystals, C1 and C2 is around 121 and 53.6 μΩ cm, respectively. These values are comparable to or slightly smaller than those reported for the isostructural compound $SrPd_2Ge_2$ [6]. Above 50 K, $\rho$ exhibits an almost linear or quasi linear temperature dependence due to electron-phonon scatterings. Such a linear or quasi linear behavior of $\rho(T)$ has been reported in cuprate superconductors and strongly correlated systems with its origin ascribed to Planckian scattering rate [31,32]. However, an alternate explanation can be offered by the Bloch-Gruneisen model, which assumes the spherical Fermi surface of metal limited to the first Brillouin zone and phonon spectrum described by the Debye temperature. At low temperatures (below 35 K), we observe $\rho \propto T^2$, a key feature of Fermi-liquid behavior. The least-square fitting to the data (shown in Fig. S3) by $\rho(T) = \rho_0 + A_{e\text{-}e}T^2$ at low temperatures (10 - 35 K) yields residual resistivity $\rho_0 \approx 12.17$ μΩ cm and electron-electron scattering coefficient $A_{e\text{-}e} \approx 2.40$ nΩ cm/K$^2$, which is comparable to that of $BaCo_2As_2$ (~ 3.94 nΩ cm/K$^2$) [33]. Below 10 K, we observe a deviation from Fermi liquid behavior, marked by the emergence of a resistivity plateau as depicted in Fig. S3 and the insets of Fig. 4. A similar resistivity plateau has been reported in various 122 phase topological semimetals [9, 21]. However, unlike other 122 topological semimetals, where metallic behavior at low temperatures is fully suppressed under a magnetic field, leading to the onset of semiconducting behavior [9,17,21], we do not observe this suppression up to 10 T ($B \parallel c$, $I \parallel ab$) as shown in bottom inset of Fig. 4.

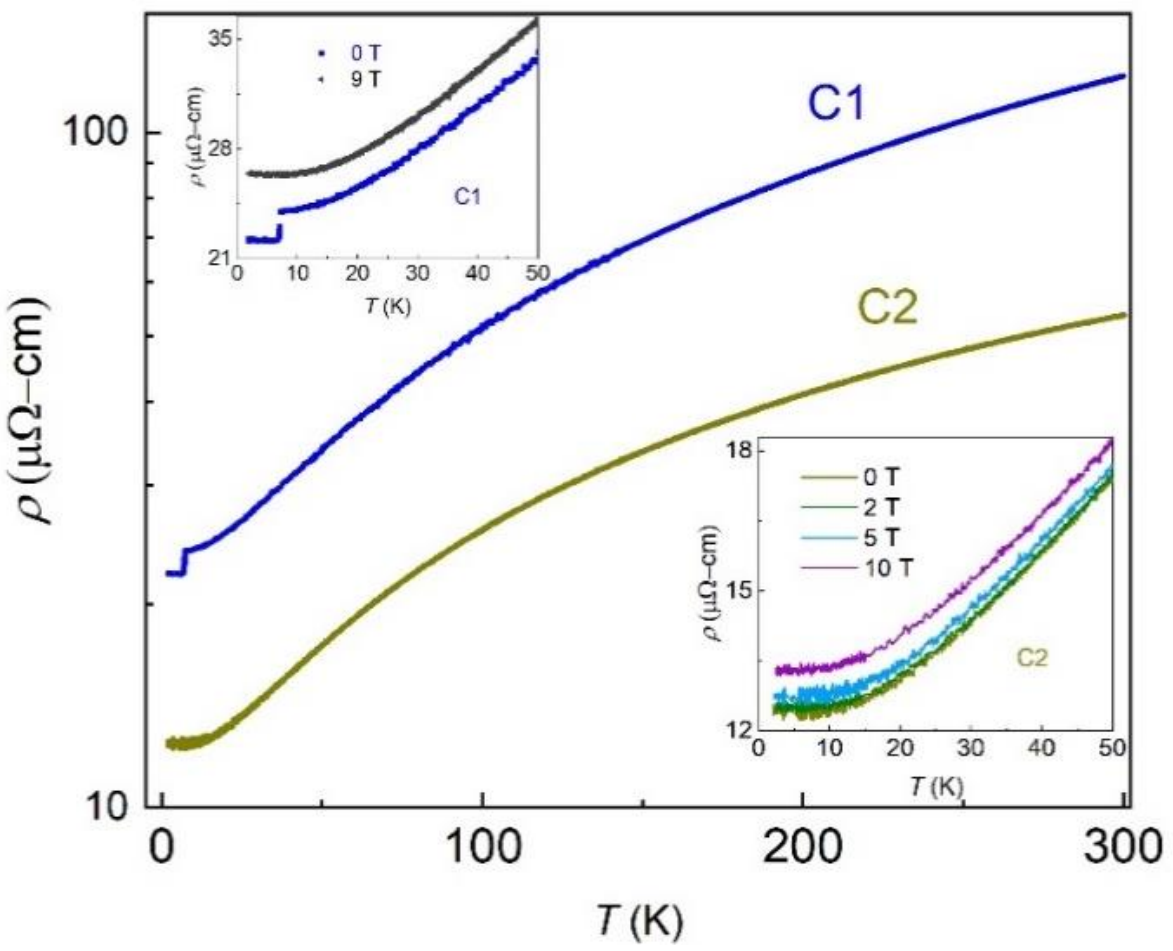

**Fig. 4.** The temperature dependence of in-plane electrical resistivity of $SrZn_2Ge_2$ crystals denoted by C1, and C2 at zero field for current applied within the $ab$-plane. The top and bottom insets show the electrical resistivities of crystals C1 and C2 respectively, under different field strengths when $B \parallel c$, and $I \parallel ab$.

### 4.4. *Magnetoresistance*

Figures 5(a) and 5(b) present the transverse and longitudinal MR measured at various temperatures for field configurations $B \parallel c$ and $B \parallel ab$ respectively with current $I$ flowing in the $ab$–plane. The MR is positive and monotonically increases with the field without any sign of saturation. However, the value of transverse (longitudinal) MR is small, just about 7% (5%) at 10 K and 9 T for $B \parallel c$ ($B \parallel ab$). Moreover, this value is larger than the typically reported values of 0.8 – 3 % for isostructural compounds like $SrPd_2Ge_2$ and $BaNi_2P_2$ [34,35]. MR continuously decreases with increasing temperature and becomes less than 1 % at 300 K. At low temperatures ($T \leq 50$ K), MR for both field configurations is primarily governed by the linear field-dependence term, accompanied by a small quadratic term (MR = a$B$ + O($B^2$)) in the high-field region, which reduces to a quadratic-like response in the low-field region. This type of crossover behavior has been reported in several Dirac materials [36-41]. To further confirm the existence of a linear MR component in data, we performed a power law fit (MR $\propto B^m$) across a broad field range of 2-9 T (see supplementary Fig. S4). At higher temperatures ($T \geq 200$ K), the fitting yields the value of exponent, $m = 2$ (Fig. 5(c)). Remarkably, the latter decreases consistently with decreasing temperature, eventually approaching value of ~1.35 at 2

K. These findings clearly demonstrate that linear MR is the dominant term at low temperatures and higher fields.

The transition from quadratic field dependence of MR in the low-field regime to linear field dependence of MR in the high-field regime could be understood by considering the crossover (at a certain critical field strength $B^*$) between the semiclassical and quantum transport regimes. According to Abrikosov, the quantum limit could be achieved if all the carriers occupy the lowest Landau level (LL) and splitting between LLs ($\Delta_{LL}$) is larger than the Fermi energy $E_F$ of the system and inaccessible to the thermal fluctuations ($k_BT$) [42]. In the case of parabolic dispersive bands, the condition $\Delta_{LL} > k_BT$ is very difficult to achieve due to very high field requirements, as $\Delta_{LL}$ is directly proportional to $B$ ($\Delta_{LL} = \hbar eB/m^*$). However, in the case of Dirac fermions, the above-described condition could be achieved at moderate field strengths because of the square root $B$-dependence of $\Delta_{LL}$ ($= \pm v_F\sqrt{2e\hbar B}$). Although we do not reach the absolute quantum limit up to the field strength of 9 T (as evident from the absence of Shubnikov de Haas oscillations), but even at this field strength, linear quantum MR could be observed with more than one LL filled [36]. The Abrikosov's quantum limit can be readily achieved in systems doped with light impurities or those characterized by a small Fermi surface and a low effective mass [36,42]. In Abrikosov's model, combination of disorder and linear dispersion leads to a linear MR [42-45]. However, both classical and quantum models require an inhomogeneous media to generate linear MR [36, 42-49]. The linear MR in these models is ascribed to Hall field contribution arising from macroscopically distorted current paths caused by inhomogeneities present within the system [46,48,50]. These inhomogeneities result in significant spatial fluctuations in conductivity tensor, causing corresponding fluctuations in mobilities or carrier densities and, consequently, linear MR [46,51-53]. Both Abrikosov's quantum model and classical models such as Parish-Littlewood, have been successfully applied to explain transverse linear MR in several conventional and Dirac materials like silver chalcogenides, InSb, $BaFe_2As_2$, $SrMnBi_2$, and $Cd_3As_2$ [36,37,40,45,51,53]. However, a notable limitation of these models is the omission of a discussion on longitudinal linear MR. In fact, compounds like silver chalcogenides and InSb exhibit negative and nonlinear longitudinal MR, which becomes positive only for extremely thin samples [36,45]. In our case, both longitudinal and transverse MR are linear as illustrated in the Fig. 5 and Fig. S5, indicating inadequacy of Abrikosov and Parish-Littlewood's models in explaining longitudinal linear MR. Additionally, accurately determining average mobilities or carrier densities in the multiband systems through the transport experiments remains a formidable challenge. For $SrZn_2Ge_2$, Hall resistivity is strictly nonlinear at higher fields, where we observed linear MR, and it is also two to three orders of magnitude smaller than magnetoresistivity, clearly ruling out the Hall field contribution.

Recently, S. Li et al. proposed a model for different types of impurities and band dispersions in the quantum limit [54]. Their model suggests that both longitudinal and transverse MR can be linear if long-range Gaussian type impurities dominate. However, fields required are very high, and do not align with our experimental data.

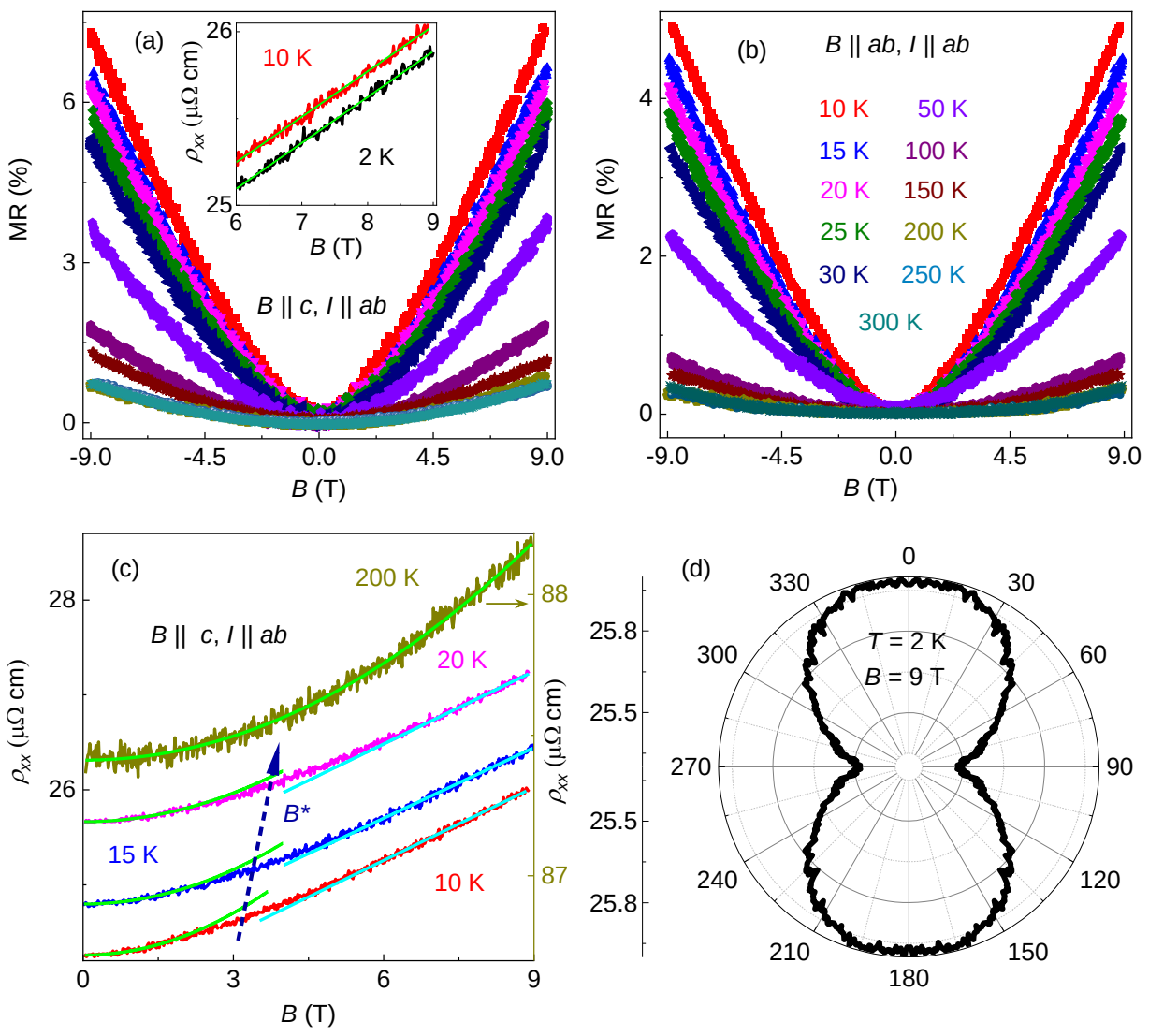

**Fig. 5.** Magnetoresistance isotherms of single-crystalline $SrZn_2Ge_2$ measured within the *ab*-plane (a) when the field is applied along the *c*-axis and (b) along the *ab*-plane. Inset in (a) shows the linear behavior of magnetoresistivity at temperatures of 2 and 10 K as a function of the field. (c) The transverse resistivity ($B \parallel c$, $I \parallel ab$) as a function of field $B$ at a few selected temperatures of 10, 15, 20 and 200 K. The green and cyan lines are quadratic and linear fits to the data respectively. The dashed line arrow illustrates the critical field $B^*$ that marks the boundary between the onset of quadratic and linear behavior of magnetoresistance. (d) Polar representation of the angular dependence of resistivity measured at $B$ = 9 T and $T$ = 2 K. θ is the angle defined between the magnetic field and the *c*-axis of the crystal.

The electronic band structure calculations for $SrZn_2Ge_2$ reveal the presence of multiple Dirac bulk and surface states near the Fermi level (see Figs. 2(d), 2(e) and Fig. 3). These states coexist with multiple conventional electron and hole pockets. Both the upper Dirac cone and nontrivial topological surface states intersect the Fermi level, as depicted in Fig. 3, suggesting their potential contribution to linear MR. Nevertheless, competitive dynamics between Dirac and conventional fermions and their role in the observed phenomena is matter of further experimental investigations including temperature-dependent angle-resolved photoelectron spectroscopy. It is noteworthy that the observed transverse linear MR in isostructural compounds like $BaFe_2As_2$ and $SrMnBi_2$ has been associated with Dirac cone states [37,40].

Furthermore, we would like to mention the other possible mechanisms of linear MR reported in the literature. One such mechanism is the formation of open orbits in the Fermi surface [55]. The open orbit is present in the hyperboloid-shaped Fermi surface of the compound (shown in the purple color in Fig. 2(j)). Nevertheless, it cannot explain linear MR for $B \parallel c$, where the magnetic field vector scans the closed Fermi surfaces in a plane normal to them. It may, however contribute to the linear MR observed for $B \parallel ab$. Nonetheless, the linear MR cannot be solely attributed to this mechanism.

Hinlopen *et al.* [56] recently proposed a specific model for linear to quadratic field dependence of MR in a wide range of materials. The key postulate of this specific model is impeded cyclotron motion on the Fermi surface, which can result in linear-*B* to quadratic crossover behavior [56]. Such an impeded cyclotron motion can occur, for example, due to the Fermi surface sectors, partially gapped Fermi surfaces, van Hove singularities, etc. [56]. Given that this model is applicable to multiband systems with any arbitrary shape of Fermi surface and that $SrZn_2Ge_2$ has multiple boundaries on the Fermi surface, it could possibly explain the MR trend in our data. The Fermi surface topology of $SrZn_2Ge_2$ is shown in the section 4.1 (see Figs. 2(f)-(k)), where the negative curvatures of dominating tetragonal, four-fold shaped Fermi surfaces are evident in moving from one vertex to another. It could result in alternating Fermi surface sectors with electron- and hole-like characters. However, some caution must be exercised in pursuing this model despite the supportive features of Fermi surface topology of the compound. This model for isotropic 2D Fermi surface with tetragonal symmetry predicts linear or quasi-linear magnetoresistivity and quasi-linear Hall resistivity in low magnetic fields, which seems consistent with our data. However, at very high fields, it predicts strictly linear Hall resistivity independent of carriers′ mass and relaxation time [56]. It strongly differs from our Hall resistivity data, which exhibits strong nonlinear field dependence at higher fields (see Fig. 6). Although, it should be noted that the maximum field applied in our data could be of intermediate strength as we do not observe the quantum oscillations at this field strength.

Next, we discuss the less popular theory of Falikov and Smith to explain linear MR [57]. This theory is based on the coherent, and anisotropic static quantum mechanical fluctuations in the ground state of electron gas in the crystal. These fluctuations originate from electron-phonon and electron-electron interactions. Interestingly, this theory predicts linear increase in both longitudinal and transverse MR without the necessity of very high magnetic fields. The role of Fermi surface anisotropy is also important here, which provide hot spots for such behavior [57]. It is interesting to note that dominating component of linear MR in our data is observed below 50 K, where we observe Fermi liquid behavior, indicative of electron-electron interactions. Although the weak linear MR is also seen at somewhat higher temperatures (see Fig. S4), indicating the important role of electron-phonon interactions also.

Further theoretical and experimental investigations are required to unravel the true mechanism behind linear MR, whether it stems from disorder, topological features or a combination of both. This necessitates the exploration of materials exhibiting linear behavior in both longitudinal and transverse MR. $SrZn_2Ge_2$ serves as a notable example in this context.

In Fig. 5(d), we show the symmetrized polar curve of angle-dependent magnetoresistivity at 2 K and 9 T. In performing the experiment, the current was confined within the *ab*-plane of the crystal, while the magnetic field was rotated from out of plane (*c*-axis) to the in-plane direction (*ab*-plane). The dumbbell-like pattern of magnetoresistivity traces the two-fold symmetry and indicates the anisotropic Fermi surface of $SrZn_2Ge_2$. The maximum of resistivity occurs at θ = $0^0$, $180^0$ ($B \parallel c$) while minimum at θ = $90^0$, $270^0$ ($B \parallel ab$). However, the overall change in angular resistivity is very small, suggesting a very small anisotropic factor, which reflects the 3D electronic transport in the compound.

### *4.5. Hall resistivity*

Next, we discuss the field dependence of Hall resistivity ($\rho_{xy}$), which is shown in Fig. 6(a). At low temperatures ($T \leq 20$ K), the $\rho_{xy}$ is positive, indicating that the holes dominate transport in the compound. Further, the $\rho_{xy}$ behavior shifts from nearly linear at low fields to nonlinear at high fields. The maximum in $\rho_{xy}$ is observed at higher fields, which, for example, appears around 5 T for 10 K data. It gradually shifts to lower fields and is suppressed as the temperature rises. Above 30 K, $\rho_{xy}$ becomes negative across the entire field range. Remarkably, the nonlinear behavior of $\rho_{xy}$ persists up to the highest measured temperature of 150 K. This is consistent with the multiband transport picture predicted by the presence of various electron and hole pockets at the Fermi level, as shown in the electronic band structure of the compound in Figs. 2(d) and 2(e).

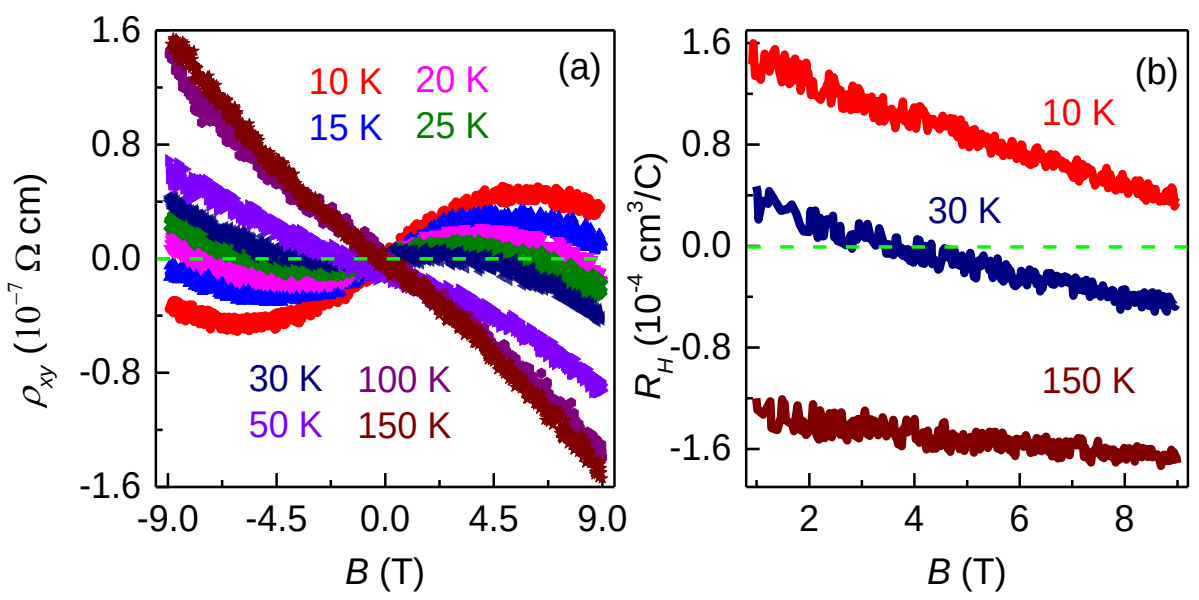

**Fig. 6.** (a) Hall resistivity of single-crystalline $SrZn_2Ge_2$ as a function of the magnetic field measured in the temperature range 10-150 K. The current is applied along *ab* – plane and the magnetic field is applied along *c*-axis of crystal. (b) The field dependence of Hall coefficients at a few selected temperatures of 10, 30, and 150 K. The low-field data points are omitted for clarity.

In the literature, for quantitative analysis of nonlinear $\rho_{xy}$, usually, a two-band model is employed. This model is suitable for multiband systems with a more or less spherical Fermi surfaces and constant and isotropic relaxation times within the bands [58]. In the case of $SrZn_2Ge_2$, the Fermi surface is quite complex and strongly deviates from the spherical Fermi surface. The isotropic relaxation time approximation is also unlikely to hold due to the very anisotropic Fermi surface. Therefore, we have calculated the Hall coefficient $R_H \approx 1.49 \times 10^{-4}$ cm$^3$/C at 10 K from the low-field region ($B < 2$ T). This value is comparable to those reported in other isostructural compounds, such as $SrPd_2Ge_2$ and $BaNi_2P_2$ [34,35]. Although, one noted difference of these compounds with $SrZn_2Ge_2$ is that they show electron carriers predominance in low temperatures.

Furthermore, we present a qualitative discussion of $\rho_{xy}$. The appearance of maximum at higher fields and negative $\rho_{xy}$ at low temperatures, which indicates carrier switching from holes to electrons, could be understood by considering the different temperatures and field-dependences of the average mobilities of electrons ($\mu_e$) and holes ($\mu_h$), with the former ones being more mobile. As the temperature increases, both $\mu_h$ and $\mu_e$ decrease, but $\mu_e$ remains relatively larger than $\mu_h$. This results in a negative $R_H$ despite the compound having a comparatively higher carrier density of holes than electrons near the Fermi level. The variation of $R_H$ with magnetic field and temperature (Figs. 6(a) and (b)) clearly indicates the role of Fermi surface′s specific shape and geometry in the electronic transport. The in-plane view of the Fermi surface is presented in Fig. 2(g)), which shows several ovals- and tetragonal-shaped Fermi surfaces, with the latter having positive and negative curvatures of different arc lengths. It is possibly the local curvatures of the latter Fermi surfaces that give rise to anisotropy in effective masses of carriers. High mobility of electrons may occur due to relatively smaller effective masses and longer relaxation times of electrons than holes. The Dirac fermions could account for some of these relatively high-mobility electrons. This potential scenario has been suggested in multiband systems such as $BaFe_2As_2$ and $Fe_{1+y}Te_{0.6}Se_{0.4}$, where the substantial rise in carrier density at low temperatures, in absence of obvious band structure change, is construed as indicative of the emergence of Dirac fermions [37,59].

Moreover, to elucidate the sign change in $R_H$ with temperature under an applied field, we refer to the semiclassical model developed by N. P. Ong for weak field limits ($\omega\tau < 1$) in 2D metals [60]. This model has been generalized for 3D systems and can be employed for multiband systems such as $SrZn_2Ge_2$. According to this model, $\rho_{xy}$ (or $\sigma_{xy}$) is determined by the Stokes area $A_l$ [= $\int dl(k)\times l(k)$] swept out by the vector $l(k)$ (= $v_k\tau_k$).

The circulation of vector $l(k)$ is directly related to the local curvature of the Fermi surface. When the circulation of $l(k)$ is different in different regions, $\rho_{xy}$ (or $\sigma_{xy}$) changes sign. In $SrZn_2Ge_2$, the dominant Fermi surface has a four-fold symmetry and comprises segments of large curvature sandwiched between the segments of small curvature. This geometry is ideal for generating different circulations of $l(k)$, leading to the sign change in $R_H$.

## 5. Conclusions

In summary, we presented detailed electronic transport properties of single-crystalline $SrZn_2Ge_2$. The electrical resistivity measured in the *ab*-plane shows $T^2$ dependence in the temperature range 10 - 35 K. Below 10 K, the compound shows a pronounced resistivity plateau, which remains robust up to the magnetic fields of 10 T. The transverse and longitudinal magnetoresistance exhibit a crossover from $B^2$ dependence in weak fields to linear-$B$ dependence in high fields at low temperatures ($T \leq 50$ K). Several topological bulk and surface states near the Fermi level, including non-trivial surface states, as evident from the electronic band structure calculations, provide evidence for the existence of Dirac fermions in $SrZn_2Ge_2$. The overall behavior of the observed magnetoresistance is understood within the framework of classical and quantum transport models, encompassing both conventional and Dirac fermions. Analysis of the temperature and field dependence of Hall resistivity, including its sign change from positive to negative at higher fields and temperatures, suggests that high mobility electron charge carriers dominate the electronic transport properties of $SrZn_2Ge_2$ despite the compound being a multiband system with multiple electron and hole pockets, as revealed by the Hall resistivity data and electronic band structure calculations.

## Acknowledgements

Z. H. acknowledges Indian Institute of Technology (IIT) Kanpur and the Department of Science and Technology (DST), India, [Grant No. DST/NM/TUE/QM06/2019 (G)] for financial support. A. C. acknowledges IIT Kanpur, and the Science and Engineering Research Board (SERB) National Postdoctoral Fellowship (PDF/2021/000346), India for financial support. A. C. and A. A. thank CC-IITK for providing the High-Performance Computing facility. P. M. acknowledges the use of Technical Research Centre (TRC) instrument facilities of S. N. Bose National Centre for Basic Sciences, India and the technical assistance of Dr. M. Roy during the measurements.

Supplementary information for "Electronic Transport and Fermi Surface Topology of Zintl Phase Compound $SrZn_2Ge_2$"

M. K. Hooda[1,2], A. Chakraborty[1,3], S. Roy[4], R. Swami[1], A. Agarwal[1], P. Mandal[5], S. N. Sarangi[6], D. Samal[6,7], V. P. S. Awana[8], and Z. Hossain[1]

[1] Department of Physics, Indian Institute of Technology, Kanpur-208016, India
[2] Department of Energy Science, Sungkyunkwan University, Suwon 16419, Republic of Korea
[3] Institute of Physics, Johannes Gutenberg Universität, Staudinger Weg 7, 55128 Mainz, Germany
[4] Vidyasagar Metropolitan College, 39, Sankar Ghosh Lane, Kolkata 700006, India
[5] Department of Condensed Matter and Material Physics, S. N. Bose National Centre for Basic Sciences, Block JD, Sector III, Salt Lake, Kolkata 700106, India
[6] Institute of Physics, Bhubaneswar, Bhubaneswar-751005, India
[7] Homi Bhabha National Institute, Anushakti Nagar, Mumbai 400085, India
[8] CSIR National Physical Laboratory, New Delhi, 110012, India

## 1. Energy dispersive X-ray (EDX) analysis of single crystalline $SrZn_2Ge_2$

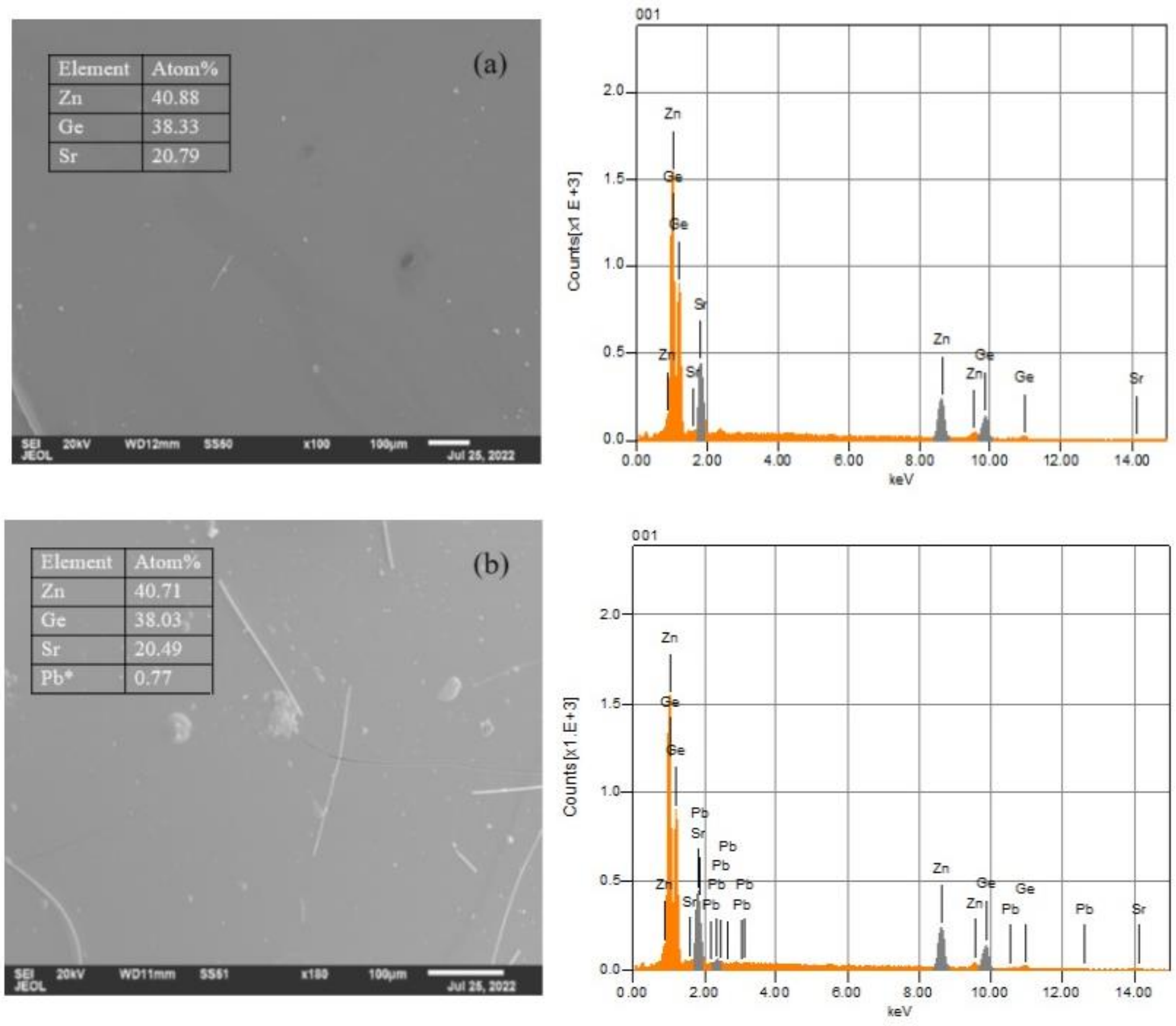

**Fig. S1**. Scanning electron micrographs and EDX spectra of single crystalline $SrZn_2Ge_2$ after centrifugation. (a) shows a clean single crystal with no detectable Pb flux residue on its surface. (b) shows the gray-white patches of minute Pb flux residue scattered randomly on the crystal surface. The star marked on Pb shows the manual selection of Pb, while performing EDX experiment on the crystal.

**Table S1- Elemental composition of $SrZn_2Ge_2$**

| $SrZn_2Ge_2$ | Atomic % | | |
|---|---|---|---|
| Sample | Sr | Zn | Ge |
| 1.1 (point) | 19.69 | 41.53 | 38.77 |
| 1.2 (point) | 20.48 | 41.30 | 38.22 |
| 1.3 (point) | 20.58 | 40.02 | 39.40 |
| 1.4 (point) | 20.89 | 41.11 | 38.00 |
| 1.5 (area) | 19.48 | 42.00 | 38.51 |
| 1.6 (area) | 20.93 | 41.66 | 37.40 |
| 1.7 (area) | 21.61 | 40.76 | 37.63 |
| 1.8 (area) | 20.98 | 40.30 | 38.72 |
| 1.9 (area) | 20.79 | 40.88 | 38.33 |
| **Average** | **20.60** | **41.06** | **38.33** |

## 2. Dirac crossings along high symmetry *X-Γ-Z-A* path

Figure S2 displays the band dispersion plotted along high symmetry *X-Γ-Z-A* path, which is presented here for better visual clarity. Several Dirac band crossings are observed around the Fermi energy. The Dirac crossings remain intact both in the non-relativistic case and with spin-orbit coupling (SOC), as shown by the red circles along the *X-Γ*. Further, SOC breaks the fourfold Dirac crossings (including spin degeneracy) around the *Z* point, resulting in two distinct Dirac crossings near the Fermi level, highlighted by the blue circles. One of the Dirac crossings is just ~ 120 meV from the Fermi energy. Additionally, SOC lifts certain band degeneracies and crossings, as shown by the blue arrow and yellow shaded region. The electronic structure closely resembles that reported in references [1-4], where the bulk band dispersion displays Dirac crossings near the Fermi energy, even in presence of SOC and shows non-trivial surface states.

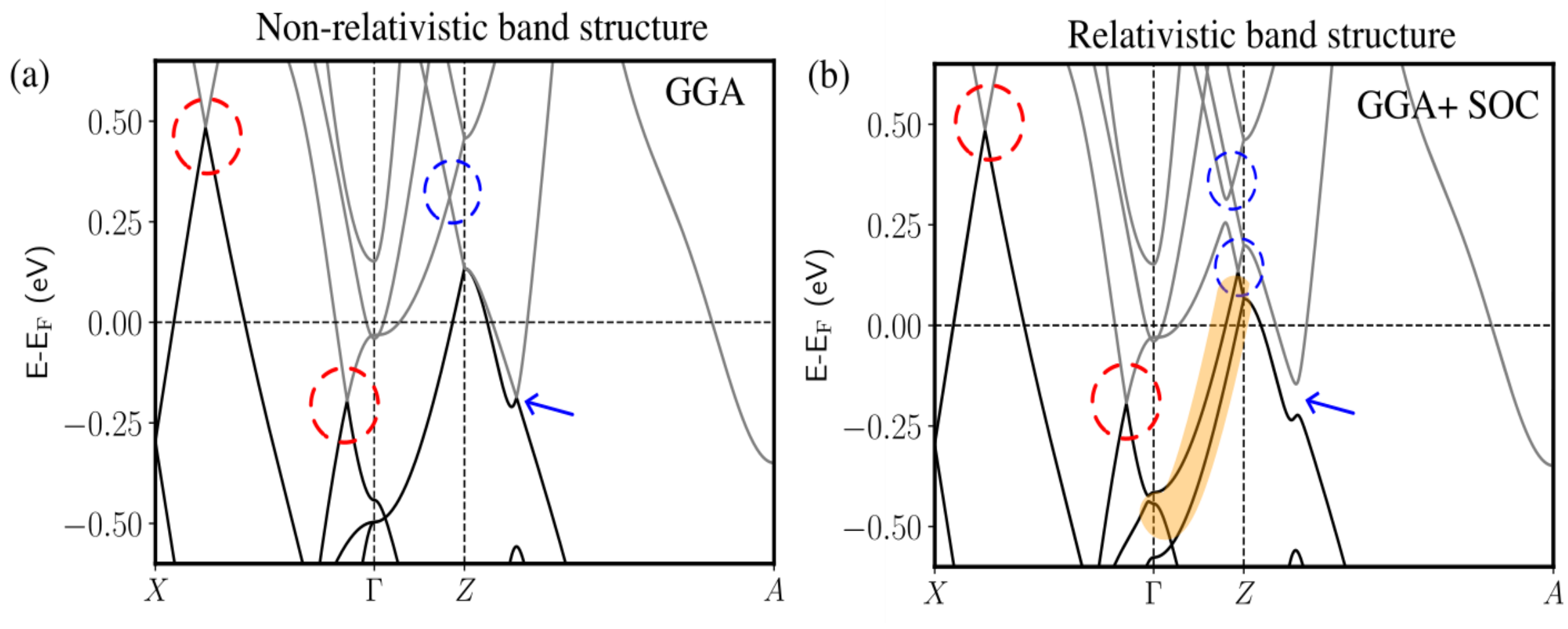

**Fig. S2**. Band dispersion along high symmetry *X-Γ-Z-A* path in absence (a) and in presence (b) of SOC. The black and grey bands represent the valence and conduction sectors respectively. The Dirac crossings are marked with red and blue dashed circles along *X-Γ-Z* path. The SOC gap openings are marked by blue arrow (away from *Z* point) and yellow shaded region (along *Γ-Z*).

## 3. Resistivity plateau and Fermi liquid behavior

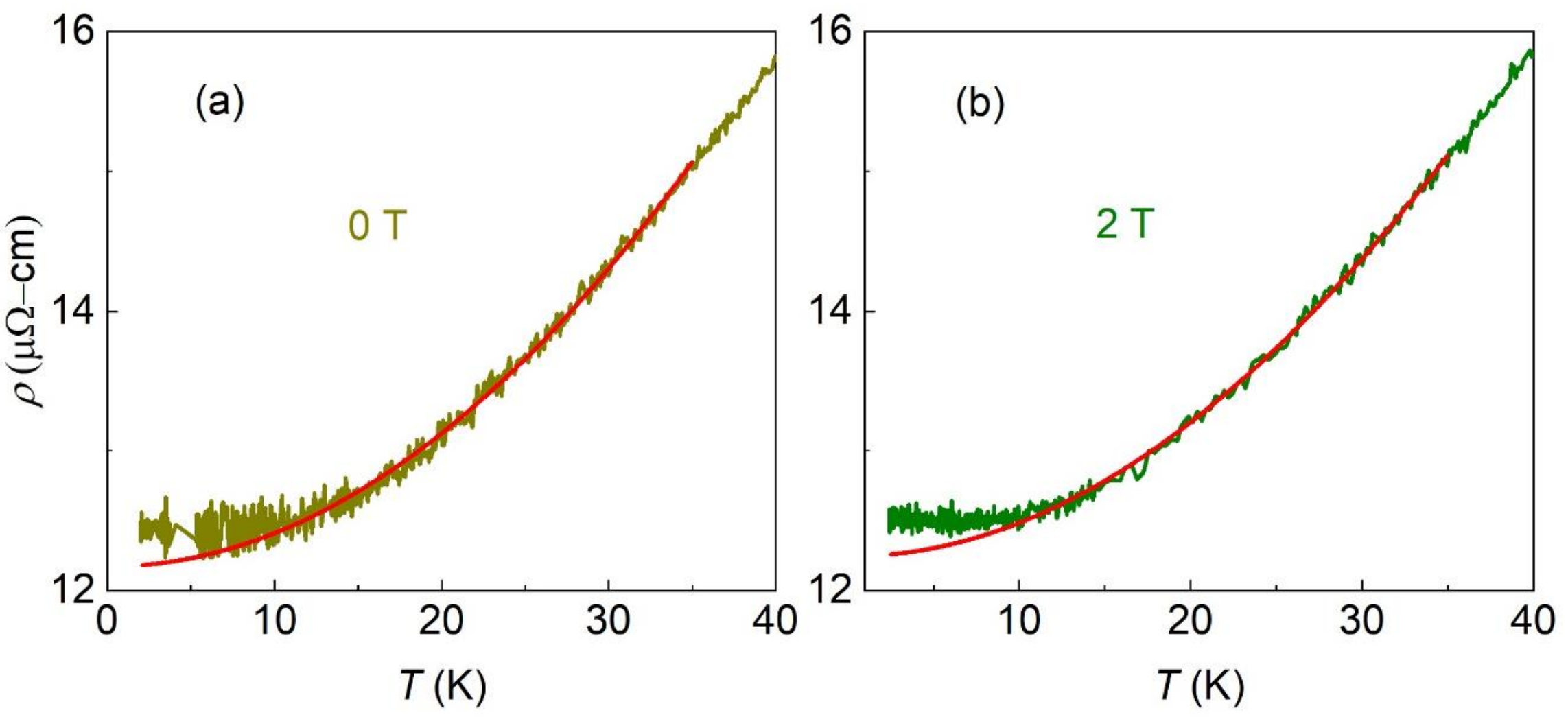

**Fig. S3**. Low temperature resistivity of crystal C2 at ***B*** = 0 and 2 T. The red line is a fit to the data using the equation $\rho = \rho_0 + A_{e-e}T^2 + A_{e-p}T^5$, where the second and third terms correspond to electron-electron (Fermi liquid) and electron-phonon scatterings, respectively. The contribution from the electron-phonon scatterings is negligible. The marked deviation from the Fermi liquid behavior below 10 K is due to the onset of resistivity plateau. A similar behavior was reported in $MoAs_2$, where a resistivity plateau emerges below 20 K, causing a deviation from Fermi liquid behavior [5].

## 4. Linear magnetoresistance at low temperatures and high magnetic fields in $SrZn_2Ge_2$

Figure S4 (a-h) illustrates the field dependence of transverse magnetoresistance (MR) with a power law fit (MR $\propto B^m$) in the field range of 2-9 T. The exponent $m$, obtained from the least-square fit, as a function of temperature, is presented in Fig. 4(i). For $T \geq 200$ K, the exponent $m$ value is 2. However, with decreasing temperature, $m$ gradually decreases in a monotonic fashion, and tends to approach the linear behavior at very low temperatures, reaching ~ 1.35 at 2 K. A similar power-law dependence has been observed in well-known topological semimetals like NbP and TaP [6,7]. Furthermore, at fields higher than 5 T, $m$ reduces to ~ 1.09 at 2 and 10 K (see the inset of Fig. 5(a) in the main text and Fig. S5), indicating the presence of a dominant linear MR component, which coexists with a very small quadratic MR contribution. This kind of coexisting MR behavior is frequently reported in several topological semimetals [8-10].

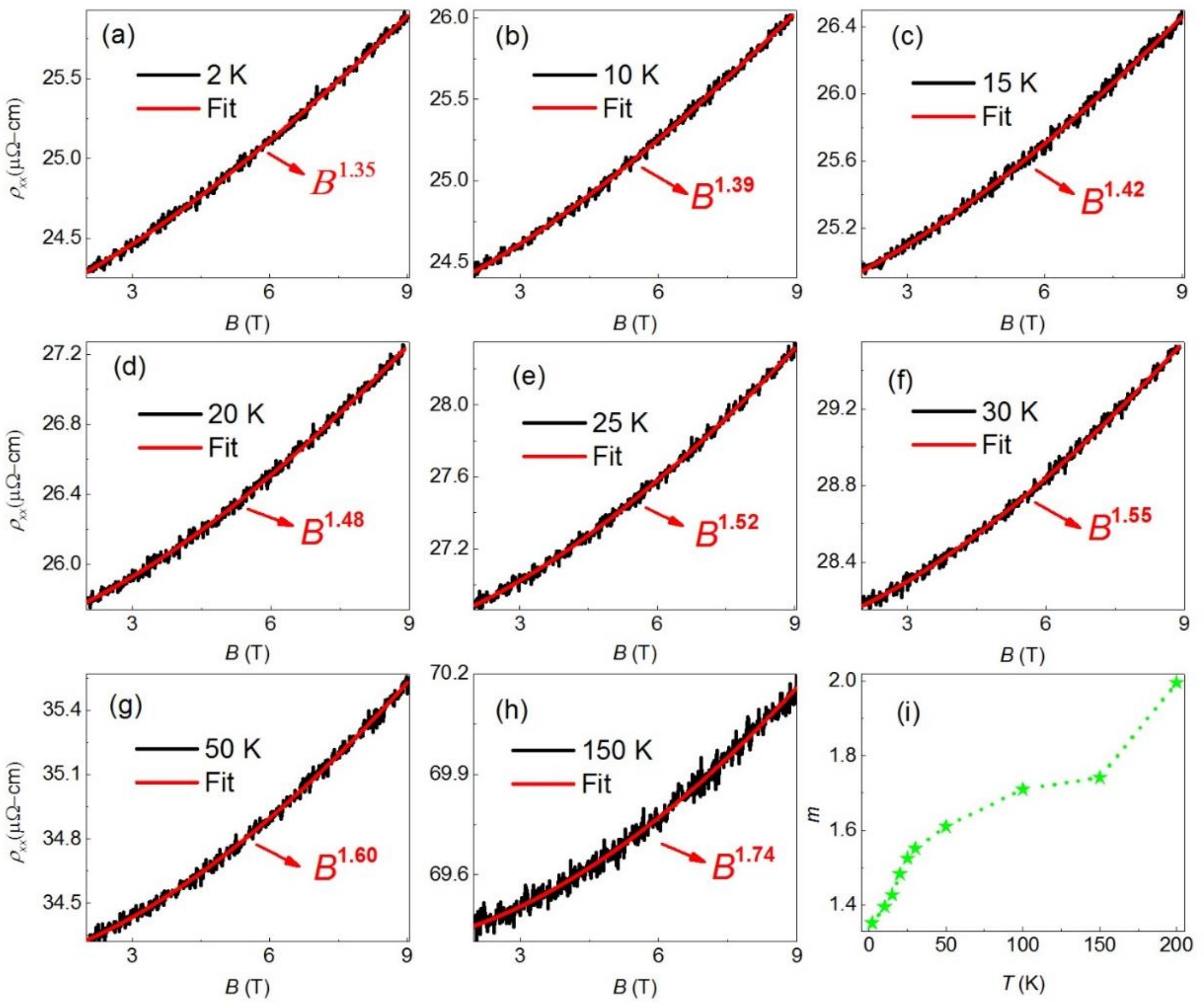

**Fig. S4.** The exponent *m* obtained from the power law (MR $\propto B^m$) fit to the transverse MR data at various temperatures in the field range 2 to 9 T. The systematic decrease in *m* value with decreasing temperature indicates the predominance of linear component of MR at lower temperatures. At higher temperatures, linear component becomes weaker and quadratic component starts dominating. At $T = 2$ K, *m* value is ~ 1.35 for $2 \leq B < 9$ T.

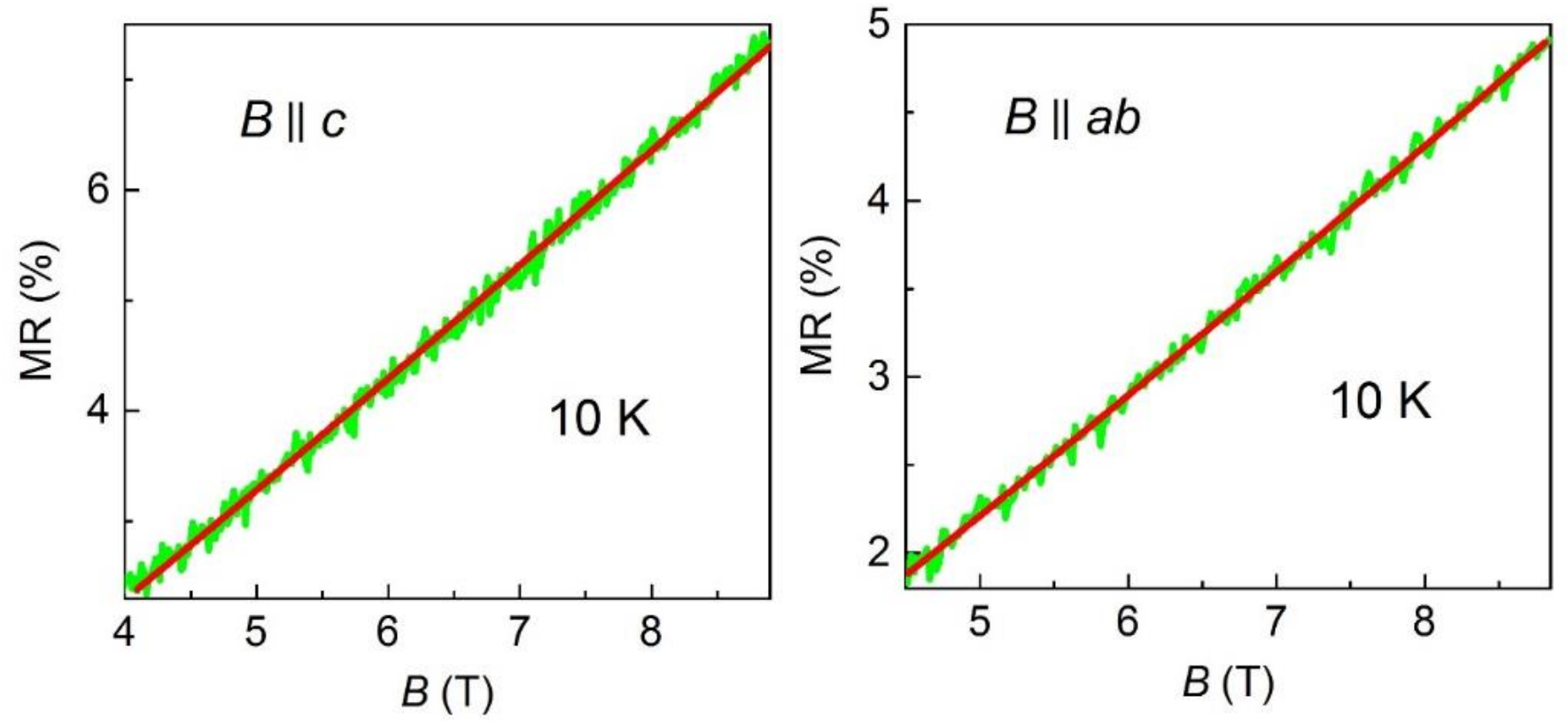

**Fig. S5.** The linear field dependence of MR of $SrZn_2Ge_2$ at temperature of 10 K for field orientations *B* ∥ *c* and *B* ∥ *ab*. The electric current is applied within the *ab* plane of the crystal. The data is already presented in the Fig. 5 of the manuscript and is shown here for better clarity. The solid red line is the guide to a linear fit to the data.